\documentclass[
superscriptaddress,
reprint,
showkeys,
 amsmath,amssymb,
 aps,
]{revtex4-1}
\usepackage{float}
\usepackage{siunitx}
\usepackage{graphicx}
\graphicspath{{Images/}}
\usepackage{dcolumn}
\usepackage{bm}
\usepackage{hyperref}
\usepackage[english]{babel}
\usepackage{pgfplots}
\usetikzlibrary{pgfplots.groupplots}
\pgfplotsset{compat=newest}
\pgfplotsset{plot coordinates/math parser=false}
\usepgfplotslibrary{fillbetween}
\usetikzlibrary{external}
\usetikzlibrary{patterns}

\usepackage{xcolor} 
\definecolor{cred}{RGB}{228,28,68}
\definecolor{cdarkred}{RGB}{129,16,38}
\definecolor{cmiddarkred}{RGB}{196,34,68}
\definecolor{clightred}{RGB}{225,100,131}
\definecolor{cbrightred}{RGB}{225,171,188}
\definecolor{cblue}{RGB}{21,102,172}
\definecolor{cdarkblue}{RGB}{14,67,113}
\definecolor{cmiddarkblue}{RGB}{0,83,152}
\definecolor{clightblue}{RGB}{31,151,255}
\definecolor{cbrightblue}{RGB}{119,184,255}
\definecolor{cgreen}{RGB}{113,171,157}
\definecolor{cdarkgreen}{RGB}{76,122,111}
\definecolor{clightgreen}{RGB}{155,195,183}
\definecolor{cviolet}{RGB}{171,55,122}
\definecolor{cdarkviolet}{RGB}{196,115,162}
\definecolor{clightviolet}{RGB}{196,115,162}

\begin{document}
\title{Comparison of Sensitivity and Low Frequency Noise Contributions in GMR and TMR Spin Valve Sensors with a Vortex State Free Layer }

\author{Herbert Weitensfelder} 
\affiliation{Christian Doppler Laboratory for Advanced Magnetic Sensing and Materials, Faculty of Physics, University of Vienna, Boltzmanngasse 5, 1090 Vienna, Austria}

\author{Hubert Brueckl} 
\affiliation{Department for Integrated Sensor Systems, Danube University Krems,  Austria}

\author{Armin Satz} 
\affiliation{Infineon Technologies Austria AG, Siemensstrasse 2, 9500 Villach, Austria}

\author{Klemens Pruegl} 
\affiliation{Infineon Technologies AG, Wernerwerkstrasse 2, 93049 Regensburg, Germany}

\author{Juergen Zimmer} 
\affiliation{Infineon Technologies AG, Am Campeon 1-12, 85579 Neubiberg, Germany}

\author{Sebastian Luber} 
\affiliation{Infineon Technologies AG, Am Campeon 1-12, 85579 Neubiberg, Germany}

\author{Wolfgang Raberg} 
\affiliation{Infineon Technologies AG, Am Campeon 1-12, 85579 Neubiberg, Germany}

\author{Claas Abert} 
\affiliation{Christian Doppler Laboratory for Advanced Magnetic Sensing and Materials, Faculty of Physics, University of Vienna, Boltzmanngasse 5, 1090 Vienna, Austria}

\author{Florian Bruckner} 
\affiliation{Christian Doppler Laboratory for Advanced Magnetic Sensing and Materials, Faculty of Physics, University of Vienna, Boltzmanngasse 5, 1090 Vienna, Austria}

\author{Anton Bachleitner-Hofmann} 
\affiliation{Christian Doppler Laboratory for Advanced Magnetic Sensing and Materials, Faculty of Physics, University of Vienna, Boltzmanngasse 5, 1090 Vienna, Austria}

\author{Roman Windl} 
\affiliation{Christian Doppler Laboratory for Advanced Magnetic Sensing and Materials, Faculty of Physics, University of Vienna, Boltzmanngasse 5, 1090 Vienna, Austria}

\author{Dieter Suess} 
\affiliation{Christian Doppler Laboratory for Advanced Magnetic Sensing and Materials, Faculty of Physics, University of Vienna, Boltzmanngasse 5, 1090 Vienna, Austria}

\begin{abstract}
Magnetoresistive spin valve sensors based on the giant- (GMR) and tunnelling- (TMR) magnetoresisitve effect with a flux-closed vortex state free layer design are compared by means of sensitivity and low frequency noise. The vortex state free layer enables high saturation fields with negligible hysteresis, making it attractive for applications with a high dynamic range. The measured GMR devices comprise lower pink noise and better linearity in resistance but are less sensitive to external magnetic fields than TMR sensors. The results show a comparable detectivity at low frequencies and a better performance of the TMR minimum detectable field at frequencies in the white noise limit.
\end{abstract}

\keywords{magnetoresistance; GMR; TMR; vortex; noise; sensitivity; detectivity}
\maketitle

\section{Introduction}
Magnetoresistive sensors have a great significance in industry. They are utilized in a vast spectrum of applications such as  compass-, biomedical-, automotive- and aerospace applications \cite{LUONG17, VALADEIRO17, SCHERNER14, HARTMANN11}. 
The most prominent sensor design is the spin valve principle with a basic common structure of two adjacent magnetic layers separated by a spacer layer. 
One magnetic layer has a pinned magnetisation (further referred to as pinned layer) whereas the other one changes its total magnetisation in external magnetic fields (further referred to as free layer). The electric resistance depends on the relative angle of magnetisation between the magnetic layers. \\
Recently, a new sensor concept with a circular shaped free layer was suggested \cite{VPAT15}. Due to the shape, material properties and a certain relation of thickness to diameter a flux-closed vortex magnetisation state forms when small or no external fields are applied.
Magnetoresistive spin valve sensors operated in a magnetic vortex configuration show negligible hysteresis while saturating at high magnetic field \cite{WURFT17}. 
These properties make them very attractive for linear current sensors or position sensors with the need for a high dynamic range and accuracy, since the latter is limited by hysteresis \cite{XIE17,WINDL17}.
Other examples are wheel speed or angle sensing applications \cite{GRANIG07, KAPSER08} where this principle could facilitate the mechanical adjustment as well as enabling more freedom setting the working point. \\
This paper focuses on the comparison of giant magnetoresistive (GMR) and tunnelling magnetoresistive (TMR) devices exploiting the vortex free layer design. 
The devices consist of a similar element arrangement and magnetic layer properties to study the sensitivity and low frequency noise. These parameters are determining the detectivity, a frequency dependent parameter giving the minimum detectable field to compare the measured devices.

\section{Theoretical Model}
\subsection{Noise Model}
Thermal noise is present in any dissipative structure arising from the random motions of the charge carriers \cite{VASILESCU06}. The thermal noise term can be described as frequency independent voltage noise power $S_V^2$ with the Nyquist formula containing the temperature in Kelvin, Boltzmann's constant $k_B$ and the total resistance $R$:
\begin{equation}
\label{eq:thermalnoise} 
S_{V,th}^2=4k_BTR \left[ \frac{\si{\volt^2}}{\si{\hertz}} \right]
\end{equation}
As soon as a DC current is applied on a resistive medium, $1/f$-noise, also known as pink noise, arises \cite{VASILESCU06, Hooge94, DHAGAT05, JIANG04}. In case of a magnetoresistive (MR) element, pink noise is described as the result of interactions from the charge carriers with defects and domain fluctuations in the magnetic layers. The pink noise term decreases with the frequency $f$ and is proportional to the squared current $I$:
\begin{equation}
\label{eq:pinknoise} 
S_{V,1/f}^2=\frac{\alpha (I\times R)^2}{f}=\frac{\alpha V^2}{f} \left[ \frac{\si{\volt^2}}{\si{\hertz}} \right].
\end{equation}
This noise arises from a statistical process, meaning the factor $\alpha $ from \eqref{eq:pinknoise} is decreasing with increasing sensor area \cite{LEI11, NOR02, SMITH96, EGELHOFF09, REED01}. Therefore it can be splitted into $ \tilde{\alpha} / NA$, with $A$  being the area of one sensitive element and N the number of elements connected in series.
Since both noise parts are arising from different physical processes and are therefore uncorrelated, the contributions can be summed up to the total frequency dependent noise power for the theoretical description of the GMR sensor:
\begin{eqnarray}
S_{V}^2 	& = & S_{V,th}^2+S_{V,1/f}^2 \label{eq:totGMR1}  \\
	 	& = & 4k_BTR+\frac{\tilde{\alpha}}{NA}\frac{V^2}{f} \left[ \frac{\si{\volt^2}}{\si{\hertz}} \right] \label{eq:totGMR2} 
\end{eqnarray}
In case of a biased TMR sensor an additional shot noise term has to be taken into account. 
Shot noise arises because the charge carriers are quantised at a potential barrier, leading to random arrival times \cite{BOGGS95,VASILESCU06}. 
The thermal assisted barrier crossing is therefore superimposed by a field-assisted barrier crossing of the charge carriers \cite{KLAASSEN04}. 
It is commonly expressed as a combination of Nyquist- and shot noise \cite{NOWAK99, EGELHOFF09, ISHIHARA01} giving the total noise of magnetic tunnel junctions (MTJ) connected in series:
\begin{equation}
\label{eq:totTMR} 
S_{V}^2=\frac{2eVR}{N} \coth\left( \frac{eV}{2Nk_BT}\right) +\frac{\tilde{\alpha}}{NA}\frac{V^2}{f}   \ \ \left[ \frac{\si{\volt^2}}{\si{\hertz}} \right]
\end{equation}
The first term results in Nyquist formula if $eV\ll k_BT$ or full shot noise of $2eVR$ for the case $eV \gg k_BT$, taking into account the serially connected devices \cite{NOWAK99,GUERRERO09,EGELHOFF09}. 
This term may not be suitable for all TMR structures and therefore a Fano factor is often introduced \cite{LEI11} to compensate deviations. 
Due to frequency limitations of the measurements to \SI{10}{\kilo\hertz}, the shot noise limit is not visible. 
Therefore the validity of equation \eqref{eq:totTMR} is assumed for all TMR noise calculations.

\subsection{Transfer Curve and Sensitivity}
\begin{figure}
	\centering
	\includegraphics[width=\columnwidth, keepaspectratio]{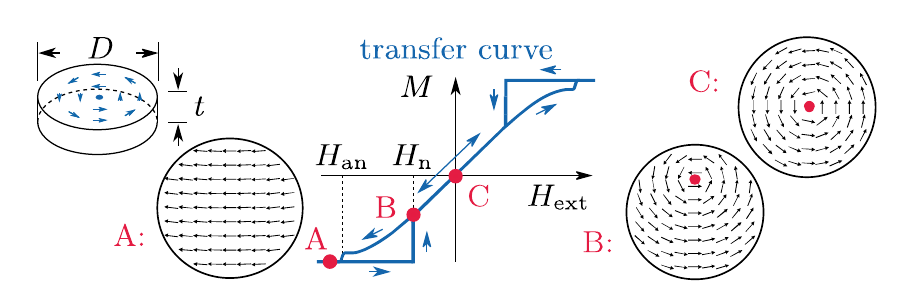}
	\caption{Theoretical transfer curve of a vortex state free layer with diameter $D$ and thickness $t$. The magnetisation M of the free layer is shown as arrows for A: in saturation, B: at the nucleation point $H_n$ and C: with no external field applied. } 
	\label{fig:transfercurve}
\end{figure}
In contrast to anisotropy pinned free layer designs, the transfer curve of vortex-state free layer sensors is characterized by a nucleation ($H_n$) and annihilation ($H_\mathrm{an}$) field (see Fig. \ref{fig:transfercurve}). 
Annihilation arises from the extinction of the vortex core in strong magnetic fields to increase the average magnetisation component towards the external field, stabilizing in a single domain state. 
This is the sensors saturation state, present until the external field decreases to the nucleation field where a stable vortex magnetisation with negligible hysteresis is formed.\\
The hysteretic behaviour between nucleation and annihilation point as well as its dependence on the material properties is studied in detail by Wurft et al. \cite{WURFT17} using micromagnetic simulations. \\
The analytical ``rigid'' vortex model, extensively described in \cite{GUSLIENKO01, GUSLIENKO01b, GUSLIENKO08}, assumes a fixed circular vortex spin structure which shifts in respect to the disc center when magnetic fields are applied. 
This shift of the vortex core leads to an increasing area where the magnetisation of the free layer is either parallel or antiparallel to  the magnetisation of the pinned layer, depending on the direction of the external field. 
The movement of the vortex center as a response to external magnetic fields is isotropic for the lateral dimensions \cite{GUSLIENKO01}, but in this context external applied fields are aligned parallel or antiparallel with the magnetisation of the pinned layer.\\
The MR effect is defined in spin valve systems as
\begin{equation}
\label{eq:xMReffect} 
\mathrm{MR}=\frac{R_\mathrm{max}-R_\mathrm{min}}{R_\mathrm{min}}\times 100\text{  }[\si{\percent}].
\end{equation}
The sensitivity is related to the magnetic stiffness of the free layer, dependent on the aspect ratio $\beta=t/D$ with the thickness $t$ and the diameter $D$ of the vortex state free layer.
The  normalised sensitivity $\gamma_{R,G}$  is calculated by the derivative of the resistance or conductance with respect to the change in applied field strength:
\begin{equation}
\label{eq:sensitivity} 
\gamma_R=\frac{\Delta R}{\Delta B}/R_0 \left[ \frac{1}{\si{\tesla}}\right] ;  
\gamma_G=\frac{\Delta G}{\Delta B}/G_0 \left[ \frac{1}{\si{\tesla}}\right].
\end{equation}
$R_0$ is the resistance and $G_0$ the conductance of the sensor at zero external field.\\
In TMR devices, the conductance is in first order linear with the cosine of the relative angle between the pinned and free layer magnetisations.\\
The detectivity $D$ describes the minimum detectable field \cite{GUERRERO09, STUTZKE05} and is given as the sensors voltage noise divided by the voltage dependent sensitivity. It is commonly used to compare different sensors:
\begin{equation}
\label{eq:detectivity} 
D=\frac{\sqrt{S_V^2}}{\gamma_R \times V} \left[ \frac{\si{\tesla}}{\sqrt{\si{\hertz}}}\right].
\end{equation}

\section{Sample Preparation and Measurement Set-Up}\label{sec:sampleprep}
The pinned layers consist of a physical vapour deposited (PVD) antiferromagnetically coupled CoFe/Ru/CoFe stack with a PtMn pinning layer. In case of GMR, copper is used as spacer layer, TMR has a MgO barrier instead.
In both cases the free layer has a cylindrical shape pattered by photolithography and ion beam etching. 
It has a thickness of \SI{80}{\nano\metre} and is made of a cobalt-iron based alloy. 
All samples consist of an array with ten elements connected in series.
In GMR the current is in plane with the element contacted at the bottom. 
In TMR the current is perpendicular to plane with the contacts above and below the magnetic layers.\\
The resistance measurements were performed in a calibrated electromagnet. 
For the noise measurements the DUT is supplied by a battery. 
An ultra low noise amplifier \cite{SCANDURRA11} is used with a lock-in amplifier to obtain the frequency resolved noise amplitudes. 
To avoid destruction through Joule heating or a breakdown of the barrier, the sensors were biased with not more than \SI{37}{\milli\volt}({\SI{500}{\micro\ampere}) in case of GMR or a junction voltage of \SI{300}{\milli\volt_j} for TMR respectively. 

\section{Measurements and Results}
\subsection{Field Measurements}
\begin{table}[tbp]
\centering
\footnotesize
\begin{tabular}{lllllll}
 & & D & $R_\mathrm{0,tot}$ &  $R_\mathrm{min,tot}$  &  RAP & MR \\
 & & [\si{\micro\metre}] & [\si{\ohm}] & [\si{\ohm}] & [\si{\kilo\ohm\micro\metre^{2}}] & [\si{\percent}] \\
\hline
 & & & & & & \\
TMR & & 2 & 5750  & 4805  & $1.51$ & 85  \\
TMR & & 1 & 26500 & 21633 & $1.70$ & 73 \\
GMR & & 2 & 71.5  & 70.1  &  & 4.3  \\
GMR & & 1 & 68.2  & 66.9  &  & 3.9  \\
\end{tabular}
\caption{Properties of the sensors consisting of 10 elements connected in series with a free layer thickness of \SI{80}{\nano\metre}. MR is the GMR- or TMR effect respectively, TMR values are given for a bias voltage per junction of $V_j=$\SI{10}{\milli\volt}. A detailed voltage dependent TMR analysis is given in figure \ref{fig:TMReffect}.}
\label{tab:sensorproperties}
\end{table}
TMR and GMR samples with different free layer diameters have been investigated.
The resistance parameters are given in Table \ref{tab:sensorproperties}, the transfer characteristic for all sensors are shown in Figure \ref{fig:GMR-TMR-S}. \\
Hysteresis is only present if the annihilation field of the vortex core is reached and a single domain state of the free layer is forming (Figure \ref{fig:GMR-TMR-S} major loops). 
If the vortex is present, i.e. the field remains smaller than the annihilation field, hysteresis is negligible (Figure \ref{fig:GMR-TMR-S} minor loops).
The ``rigid'' vortex model predicts a decreasing sensitivity with increasing aspect-ratio $\beta=t/D$. 
As can be seen in the measurements, a doubling of the diameter $D$ decreases $H_\mathrm{an}$, and therefore, increases the sensitivity $\gamma_{R,G}$.
\subsection{Noise Measurements}
The frequency dependent noise measurements are shown in Figure \ref{fig:noiseTMR} and \ref{fig:noiseGMR} for devices with \SI{1}{\micro\metre} and  \SI{2}{\micro\metre} free layer diameter. 
The measured signal is clearly above the noise floor of the ULNA and induced disturbances (black dashed lines in Fig. \ref{fig:noiseGMR}). 
The theoretical curves (colored lines) are fitted to the measurements (dots) with equations \eqref{eq:totGMR2} and \eqref{eq:totTMR} for GMR and TMR sensors, respectively. 
The fit includes the amplifier noise, the thermal and shot noise level as well as the pink noise with the empirical noise parameter $\tilde{\alpha}$, which is the only fit parameter. 
Bumps observed in the TMR noise spectra, leading to a deviation of the fit curves to the measurement, may be attributed to random telegraph noise (RTN) as also reported in \cite{ALMEIDA08}. 
As extracted from the fitted curves, the TMR samples exhibit a noise parameter $\tilde{\alpha}$ higher by a factor of 300 compared to the GMR samples. 
The high pink noise contribution in TMR is shifting the corner frequency (between $1/f$ and white noise floor) out of the measurement range. 
Therefore only pink noise contributions are visible. 
Although the noise parameter  $\tilde{\alpha}$ is normalised to the sensors area, it is roughly a factor of two between diameter of \SI{1}{\micro\metre} and \SI{2}{\micro\metre} for both GMR and TMR devices. 
This is attributed to the magnetic stiffness of the free layer, increasing with the parameter $\beta$ as can also be seen in the annihilation field $H_\mathrm{an}$.\\
As the TMR is decreasing with higher supply voltage, the noise parameter $\tilde{\alpha}$ is reduced as well. 
This behaviour of the noise parameter is also discussed in \cite{ALMEIDA08,HAN10}. 
Figure \ref{fig:TMReffect} shows the bias voltage dependence of the TMR and $\tilde{\alpha}$. 
Trends for the correlation of the noise parameter $\tilde{\alpha}$ with the resistance area product and the TMR effect are given in \cite{LEI11} and \cite{GOKCE06}. The values from the literature are in good agreement with the measurements. 
As this effect is not present in GMR sensors, the noise parameter is constant. 
\begin{figure}
	\centering
	\includegraphics[width=\columnwidth, keepaspectratio]{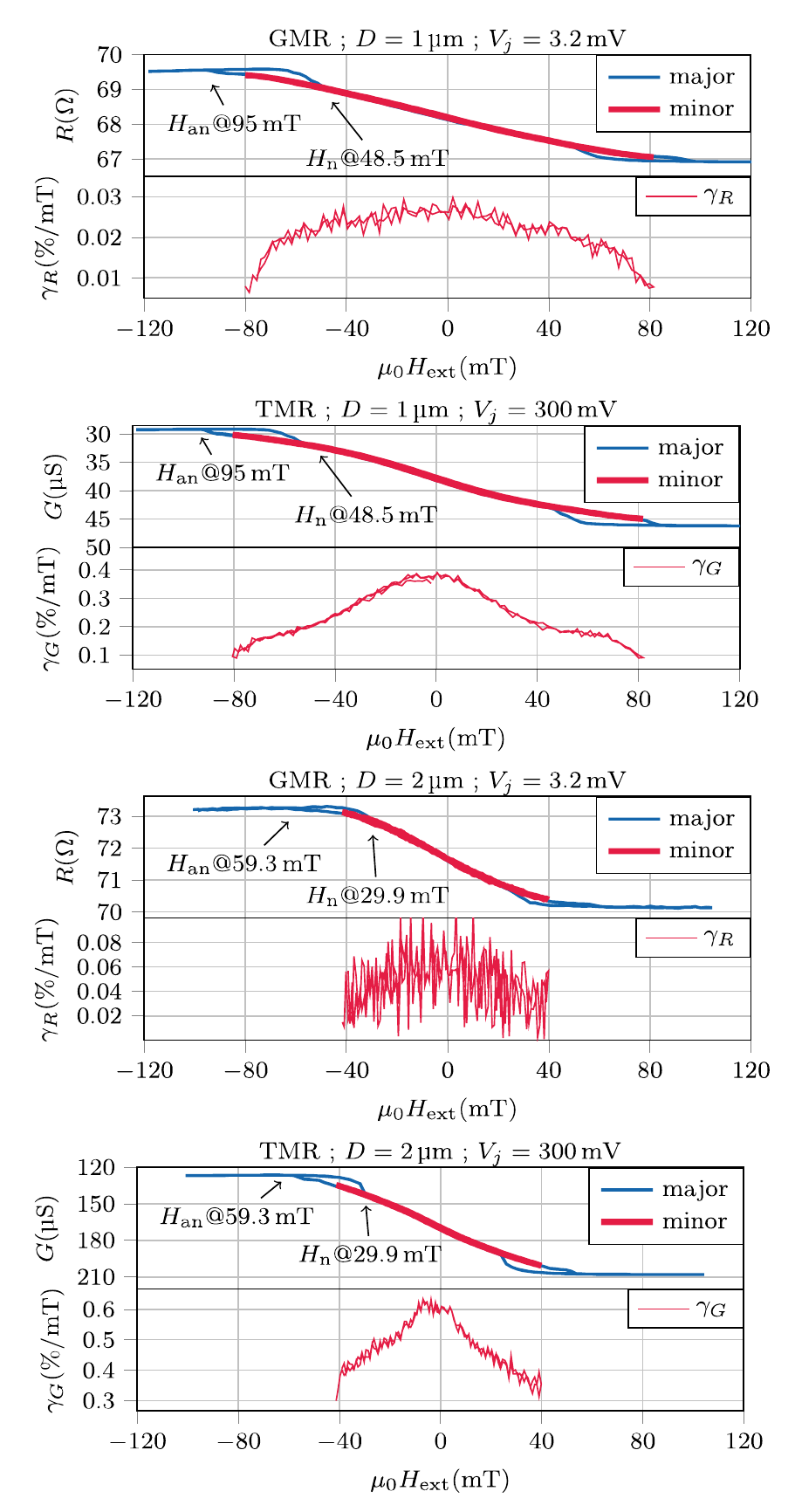}
	\caption{Comparison of the transfer characteristic of GMR and TMR samples with diameters $D$ of 1 and 2 \si{\micro\metre}. Shown are major loops ($H_\mathrm{ext}>H_\mathrm{an}$) and minor loops ($H_\mathrm{ext}<H_\mathrm{an}$) as well as the sensitivity of resistance ($\gamma_R$) and the sensitivity of conductance ($\gamma_G$) calculated using equation \eqref{eq:sensitivity}. }
	\label{fig:GMR-TMR-S}
\end{figure}
\begin{figure}
	\centering
	\includegraphics[width=\columnwidth, keepaspectratio]{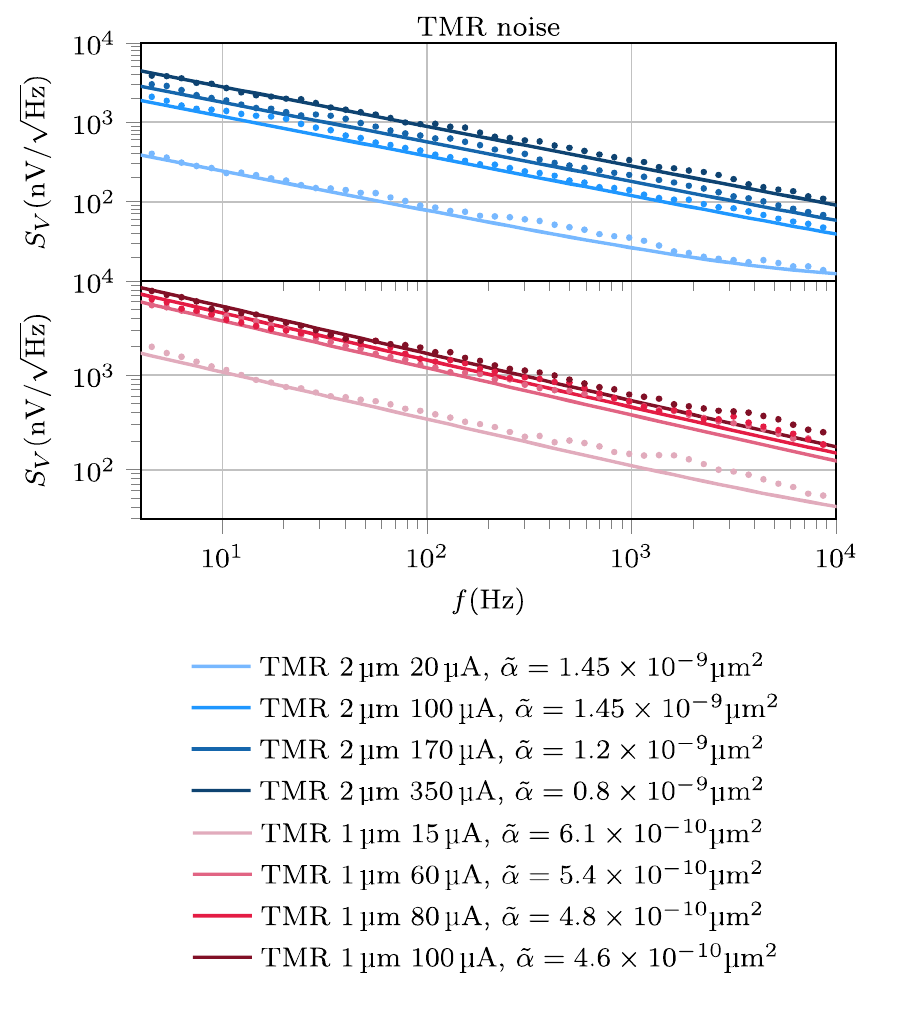}
	\caption{Noise in TMR sensors with a diameter of \SI{2}{\micro\metre} and \SI{1}{\micro\metre} at different bias currents. The measurement results (dots) are fitted (lines) according to equation \eqref{eq:totTMR} with the adjustable parameter $\tilde{\alpha}$.}
	\label{fig:noiseTMR}
\end{figure}
\begin{figure}
	\centering
	\includegraphics[width=\columnwidth, keepaspectratio]{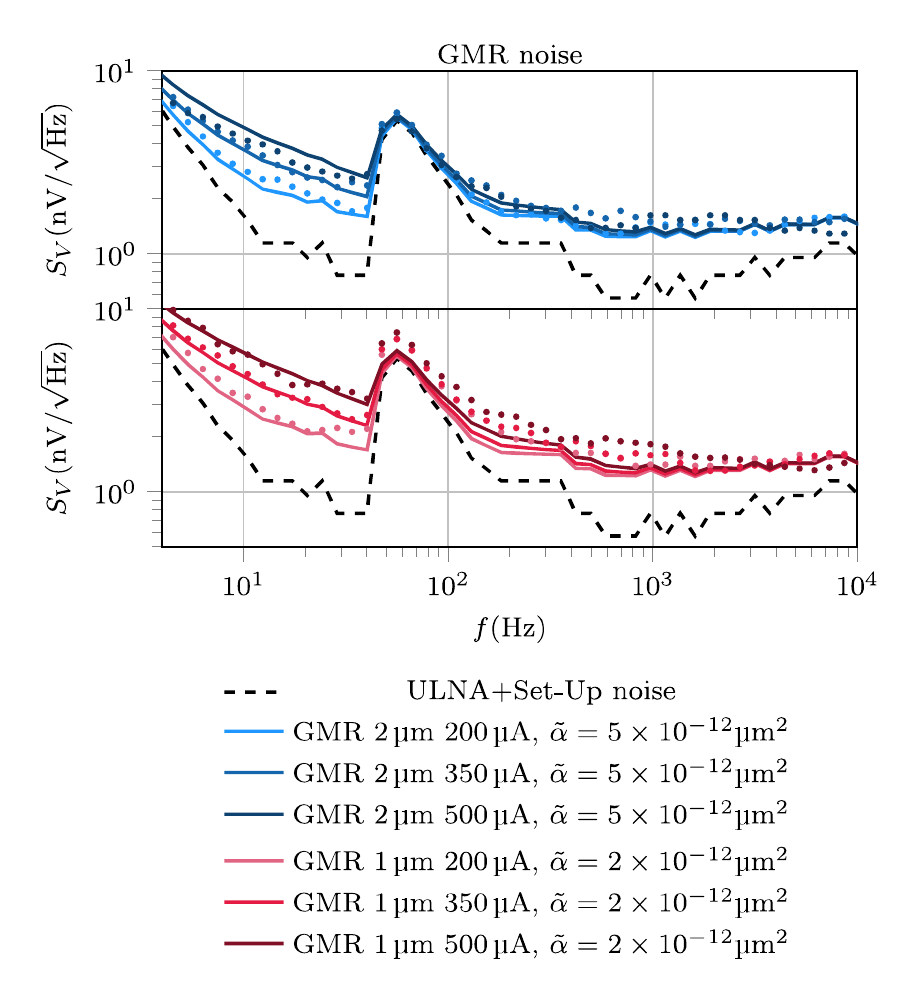}
	\caption{Noise in GMR sensors with a diameter of \SI{2}{\micro\metre} and \SI{1}{\micro\metre} at different bias currents. The measurement results (dots) are fitted (lines) according to equations \eqref{eq:totGMR2} with the adjustable parameter $\tilde{\alpha}$.}
	\label{fig:noiseGMR}
\end{figure}
\begin{figure}
	\centering
	\includegraphics[width=\columnwidth, keepaspectratio]{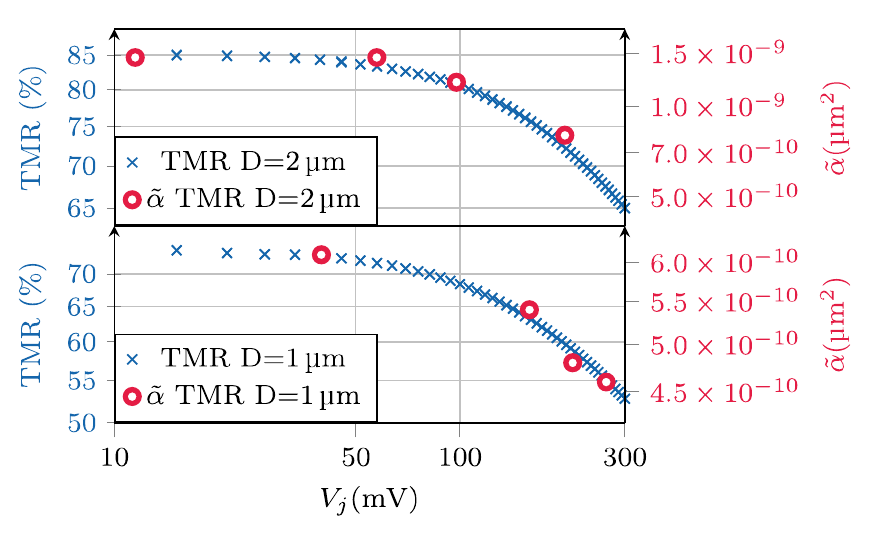}
	\caption{TMR effect of \SI{1}{\micro\metre} and \SI{2}{\micro\metre} disks (blue axis, crosses; see formula \eqref{eq:xMReffect}) with the noise parameter $\tilde{\alpha}$ (red axis, dots) extracted from the measurements in Figure \ref{fig:noiseTMR}. TMR as well as $\tilde{\alpha}$ are decreasing with the bias voltage. At low voltages $\tilde{\alpha}(D=\SI{2}{\micro\metre})$ is roughly twice of $\tilde{\alpha}(D=\SI{1}{\micro\metre})$. }
	\label{fig:TMReffect}
\end{figure}

\subsection{Detectivity Comparison}
\begin{figure}
	\centering
	\includegraphics[width=\columnwidth, keepaspectratio]{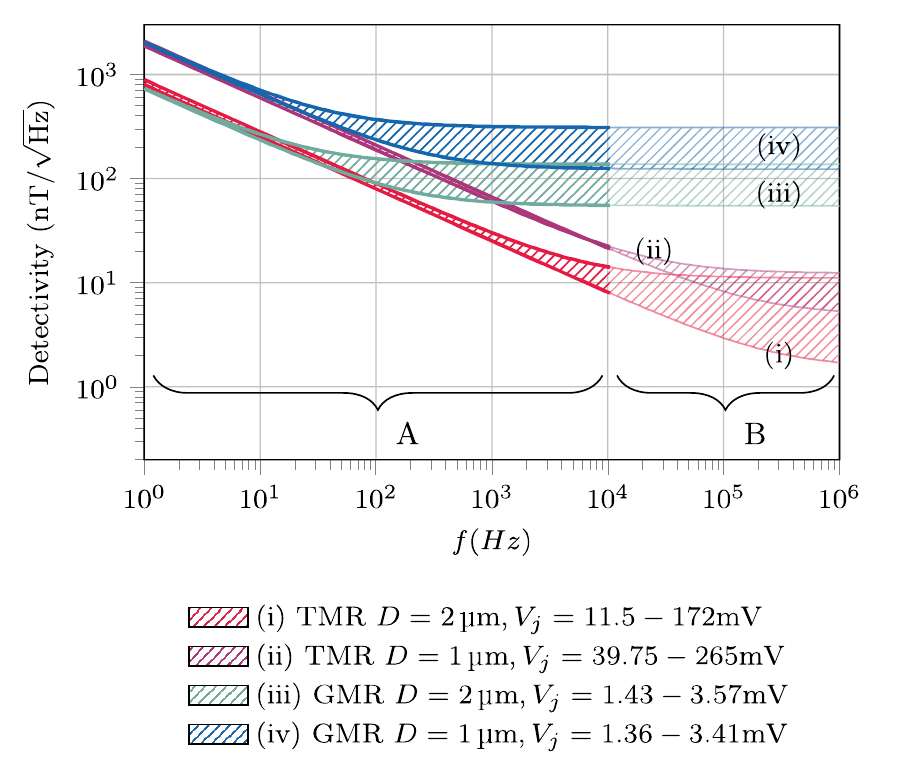}
	\caption{Calculated detectivity based on pink and white noise terms as given in equation \eqref{eq:totGMR2} and \eqref{eq:totTMR} for the measured (A) and an extrapolated (B) frequency range. The areas represent the dependence of the detectivity on the supply voltage. Higher supply voltages lead to a lower detectivity level due to the increased output voltage.}
	\label{fig:Detectivity}
\end{figure}
The detectivity comparison is shown in Figure \ref{fig:Detectivity}. 
It is calculated using equation \eqref{eq:detectivity} with the obtained parameters from the fitted noise data in Figures \ref{fig:noiseTMR} and \ref{fig:noiseGMR}. 
The plotted areas indicate the detectivity obtained from the measured array configuration of 10 serially connected devices, considering all applicable supply voltages as described in section \ref{sec:sampleprep}. 
The plot is divided into region A, where the detectivity is obtained from the fitted noise measurements, and region B, an extrapolation using the formulas \eqref{eq:totGMR2} and \eqref{eq:totTMR} for white and pink noise contributions.
A higher supply voltage is generally improving the detectivity in the high frequency regime for GMR and TMR sensors, as also observed in literature \cite{REIG13}. 
The detectivity in this regime is proportional to the sensitivity $\gamma \times V$ which is improved at a larger bias.
But since the TMR effect as well as the sensitivity and noise parameter of the TMR are decreasing with the supply voltage, the shapes differ to areas obtained by the GMR sensors. 
The same free layer diameter leads to roughly the same detectivity level in the low frequency range where flicker noise is dominant. 
Even though the noise parameter $\tilde{\alpha}_{TMR} \sim 300 \times \tilde{\alpha}_{GMR}$, the TMR sensor has the same detection limit in the low frequency regime and clearly wins in the high frequency regime considering white and pink noise contributions. 
The TMR sensors have a better detectivity performance at higher frequencies due to the better sensitivity while having only a 1.5 times larger noise floor when considering shot and thermal noise contributions as given in equation \eqref{eq:totTMR}.
The low frequency detectivity is roughly on the same level for GMR and TMR sensors with the same area. 
At \SI{10}{\hertz}, as the white noise becomes dominant, the curves are splitting up. The knee between pink and white noise contributions is dependent on design parameters like resistance, area or bias voltage. 
For practical applications, the corner frequency can be at \SI{1}{\kilo\hertz} or even higher \cite{STUTZKE05}.
One possible interpretation for the dependence of the low frequency detectivity on the sensors area is a magnetic origin of the pink noise as also stated in \cite{OZBAY06}. 
A derivation by Egelhoff et al. \cite{EGELHOFF09} for MTJ`s shows the independence of the detectivity on the magnetoresistive effect and bias voltage for pink magnetic noise.\\

\section{Conclusion and Outlook}
In this work giant- and tunnelling magnetoresistive spin valve sensors with a magnetic vortex state free layer disk were compared.
All samples were based on the same free layer thickness, free layer material and had the same amount of active elements or junctions connected in series, leading to comparable sensor properties. 
The broad linear range is given by the annihilation field of the magnetic vortex state, dependent on the ratio of thickness to diameter. 
Compared to the GMR sensors, the noise parameter $\tilde{\alpha}$ of the TMR sensors is 300 times higher.
Yet the TMR sensors show a lower detection limit at frequencies where white noise is dominant. 
This is attributed to the high TMR sensitivity.
The detectivity can be improved by using higher supply voltages, despite the fact of a thereby decreasing sensitivity in TMR sensors.
The same free layer diameter leads to roughly the same detectivity level in the low frequency range where for GMR as well as TMR sensors pink noise is dominant. 
This suggests that the noise in the low frequency regime is dominated by magnetic contributions, because magnetic pink noise is not dependent on the MR effect or the bias voltage but on the active sensor area. 
The next-stage work will be a study on segregated electric and magnetic noise contributions in vortex-state free layer devices.
\section{Acknowledgements}
The authors would like to gratefully acknowledge the Christian Doppler Laboratory Advanced Magnetic Sensing and Materials, for financial support. 
The Laboratory is financed by the Austrian Federal Ministry of Economy, Family and Youth, the National Foundation for Research, Technology and Development.

\end{document}